\documentclass[revtex4,a4paper, preprint, superscriptaddress]{revtex4}  
\usepackage[draft]{hyperref} 
\usepackage{graphicx}

\begin{document}

\title{Photodarkening of blinking quantum dots is not governed by Auger recombination}

\author{Shamir Rosen, Osip Schwartz and Dan Oron}
\affiliation{Department of Physics of Complex Systems,
Weizmann Institute of Science, Rehovot, Israel}

\begin{abstract}
The observed intermittent light emission from colloidal
semiconductor nanocrystals has long been associated with Auger
recombination assisted quenching. We test this view by observing
transient emission dynamics of CdSe/CdS/ZnS semiconductor
nanocrystals using time-resolved photon counting. The size and
intensity dependence of the observed decay dynamics are inconsistent
with the those expected from Auger processes. Moreover, the data
suggests that in the `off' state the quantum dot cycles in a
three-step process: photoexcitation, rapid trapping and subsequent
slow nonradiative decay.
\end{abstract}
\maketitle

Physical properties of chemically synthesized semiconductor
nanocrystal quantum dots (QDs) have been the subject of extensive
research in the last two decades. Utilized as fluorescent labels,
they demonstrate good photostability, high absorption cross
sections, wide excitation spectra and narrow emission lines, which
makes them an attractive alternative to organic fluorophores in a
wide range of life science applications \cite{Alivisatos_review}. In
addition, QDs can serve as tunable light absorbers and emitters in
optoelectronic devices such as light-emitting diodes and QD
sensitized solar cells.

A general property of QDs is a fluorescence intermittency apparent
in single dot emission, also known as blinking. It is observed as
abrupt jumps from a strongly emitting state to episodes of darkness
during which the emission intensity is heavily attenuated despite
continuous laser excitation \cite{nirmal}. This is an intriguing
phenomenon as it results in a clearly measurable manifestation of
microscopic dynamical changes in a single nanocrystal. Blinking is
of practical importance since it reduces the effective quantum yield
of QDs. It also limits the utility of QDs in applications such as
single particle tracking. The luminescence intensity fluctuations of
blinking QDs occur on time scales which are immensely longer than
the longest characteristic time normally associated with QD
dynamics, a radiative lifetime of tens of nanoseconds
\cite{barkai_review}. Hence, they must be associated with``slow''
variations of the microscopic state of the QD. Yet, despite the
broad literature regarding the statistics of the blinking process,
as well as its dependence on temperature, excitation wavelength and
intensity, the detailed mechanism inducing this behavior is,
surprisingly, still under debate.

The vast majority of the existing theoretical models
\cite{frantsuzov_review}, as well as much of the existing
experimental literature (including recent realizations of
nonblinking QDs
\cite{nonblinking_french,nonblinking_klimov,nonblinking_krauss,grey,mulvaney}),
associate the `off' periods with a long-lasting change in the
charging state of the QD. This may be brought about by
photoionization, as suggested originally by Efros and Rosen
\cite{efros}, or by trapping of an excited charge carrier in a
long-lived surface trap. In either case, QDs in the `off' state are
essentially ionized and can, upon further photoexcitation,
nonradiatively decay via Auger recombination. This is an intra-QD
energy transfer interaction by which the excess energy from a
recombination event is transferred to the spectator charge carrier
rather than emitted as a photon \cite{klimov_auger_dots}. In order
to account for the observed power-law statistics of `off' and `on'
times \cite{barkai_review}, several modifications of the Efros and
Rosen formulation have been recently suggested. Thus, multiple traps
with an exponential distribution of trapping and escape rates
\cite{verberk} or a fluctuating energy difference between the trap
state and the excited state \cite{bawendi_diffusion} have been
postulated.

One alternative model, which does not require the existence of
long-lived charge traps, has been suggested by Frantsuzov and Marcus
\cite{frantsuzov}. In this model, the `off' state arises from the
opening of a nonradiative decay channel of the singly excited dot.
In the proposed mechanism the excess energy from the trapping of the
hole is resonantly transferred to the electron in the lowest excited
state (1S) thereby ejecting it to the next excited state (1P). After
quickly relaxing back to the 1S state the electron recombines
nonradiatively with the trapped hole. Thus, Auger processes are not
invoked to account for photodarkening, although these should be
observed at an excitation rate exceeding the nonradiative decay
rate. The only experimental support to this model obtained so far
comes from relatively indirect statistical
measurements\cite{marcus}.

Most of the experimental work on blinking has focused on
characterization of the statistics of `on' and `off' times
\cite{barkai_review}. The study of the transient decay dynamics of
QDs during the `on' and `off' periods is a complementary approach,
which has greatly benefited from recent advances in time-resolved
photon counting instrumentation. In particular, monitoring the decay
dynamics of the remaining fluorescence during `off' periods provides
detail on the nonradiative decay mechanism responsible for
photodarkening. Such measurements revealed that the emission
transient following pulsed excitation is generally nonexponential,
and that the emission intensity is correlated with its lifetime
\cite{mews_time_res}. By creating decay curves from only the photons
arriving at periods with the highest emission rates, it was shown
that the 'on' state emission is well fit by a single exponent
\cite{bawendi_time_res}. Later work demonstrated that there exists,
in fact, a continuous distribution of emitting states
\cite{alivisatos_time_res} and found a strong correlation between
fluorescence intensity and decay times. Very recently, studies on
CdSe/CdS QDs with reduced blinking \cite{grey,mulvaney} have shown
relatively strong emission in the 'off' state. In both, this was
attributed to slower Auger dynamics due to the large QD size.

Here we attempt to elucidate the microscopic mechanism of QD
darkening by a systematic study of the `off' state dynamics in the
most studied system of CdSe/CdS/ZnS QDs. In particular, our aim is
to clarify the role of the Auger processes, which are known to be
strongly dependent on both the nanocrystal size and the excitation
intensity. We first study the `off' state fluorescence at low
illumination intensities and find that it exhibits size-independent
dynamics. This is in clear disagreement with the strong size
dependence expected from Auger-assisted photodarkening. We then
proceed to characterize the excitation intensity dependence of the
`off' state lifetimes, and discover that under strong excitation
Auger processes do dominate the nonradiative decay. Finally, we
present a phenomenological model accounting for the results, provide
guidelines for the design of non-blinking QDs, and discuss possible
experimental pathways to further elucidate the detailed dynamics of
this fundamental system.

CdSe/CdS/ZnS QDs were synthesized in three sizes following standard
procedures \cite{peng}. Briefly, QDs were grown in a
non-coordinating solvent and overcoated using successive ion
layering \cite{silar}. We used nanocrystals with diameters of 3.8nm,
5nm and 8nm (corresponding to emission peaks at 590nm, 618nm and
665nm, respectively). All QDs were slightly rodlike, with an aspect
ratio of $\approx2$. The quantum yield of the respective samples was
determined in solution to be $80\%$, $70\%$ and $15\%$. Nanomolar
concentrations of QDs in a 3$\%$ mass/volume PMMA solution were
spin-cast onto glass cover slips creating samples with typical
densities of 0.02 QDs/$\mu m^{2}$. The QDs were excited by
frequency-doubled pulses from a Ti-Sapphire oscillator at 400nm,
with a duration of 100fs and a repetition rate of 80MHz. Light was
focused on the sample by an oil immersion objective with a numerical
aperture of $1.4$. Epi-fluorescence was spectrally filtered and
detected by a single-photon avalanche photodiode (id Quantique).
Emission time traces from isolated QDs were recorded by a
time-correlated single photon counting system (Picoharp300,
Picoquant), operated in the time-tagged mode such that each photon
is assigned an absolute arrival time and an arrival time relative to
the excitation pulse. The system temporal resolution was measured to
be 65ps.

\begin{figure}
\centering
\includegraphics[width=9cm]{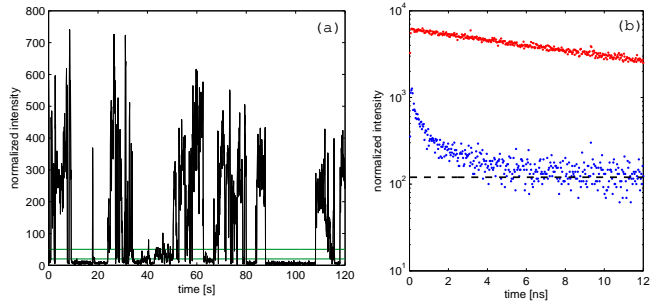}
\caption {(a) Intensity time trace of a single CdSe/CdS/ZnS QD
showing fluorescence intermittency (bin width = 60ms). The `on' and
`off' intensity thresholds are also shown. The excitation power is
adjusted to an excitation rate of about 0.2, taking into account the
radiative lifetime and absorption cross section \cite{peng_cs}. (b)
`On' (red) and `off' (blue) PL decay curves. The dotted line
represents the dark count level of the detector.} \label{timetrace}
\end{figure}

A representative intensity histogram of a blinking QD is shown in
Fig.~\ref{timetrace}(a). For each such data set we define the `on'
and `off' count rate thresholds. `On' and `off' photoluminescence
(PL) decay curves were produced by binning all photons which arrived
during the respective periods according to their time of arrival
relative to the excitation pulse. Both are presented in
Fig.~\ref{timetrace}(b). For most observed QDs the `on' state
exhibits single exponential decay, in agreement with previous
observations \cite{bawendi_time_res}. The `off' state, however,
demonstrates a more complicated decay curve with a wide range of
lifetimes. Typical `off' time decay curves contain both rapid
components of $\sim 100$ps and relatively long ones of $\sim 1$ns.
Multiexponential decays were previously observed on CdSe/ZnS QDs
\cite{bawendi_time_res}, and recently in CdSe/CdS QDs \cite{grey},
implying that such behavior is is not unique to our system.

\begin{figure}
\centering
\includegraphics[width=7cm]{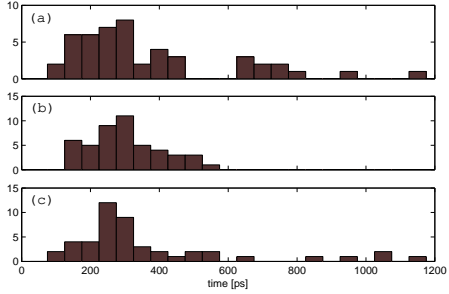}
\caption{Histograms of the 'off' state 1$/$e decay times of (a)
3.8nm QDS, (b) 5nm QDs and (c) 8nm QDs. } \label{sizes}
\end{figure}

Such measurements were performed on several tens of nanocrystals at
each of the three sizes. As we cannot directly assign a lifetime to
the `off' state decay, we resort to assigning a $1/e$ decay time to
each data set. Histograms of these decay times are presented in
Fig.~\ref{sizes} for all three sizes of QDs. The observed
distributions demonstrate a broad peak at about 250ps and are
evidently size-independent. This is in stark contrast with the
significant variance in the biexciton Auger decay lifetimes, which
are approximately 30ps, 65ps and 350ps respectively for the three
sizes \cite{klimov_auger_dots,klimov_auger_rods}. Since both the
slow and the fast components of the `off' state decay curve
contribute to the $1/e$ decay time, the lifetimes of the slower
components are significantly longer than biexciton Auger
recombination lifetimes, particularly for the smaller QDs (for
these, they are even longer than the expected trion Auger lifetime
\cite{trion}). This fact, in combination with the lack of size
dependence in the observed rates, leads us to the conclusion that
the nonradiative recombination process causing the intermittent
darkening of QDs is {\bf not} Auger recombination.

\begin{figure}
\centering
\includegraphics[width=9cm]{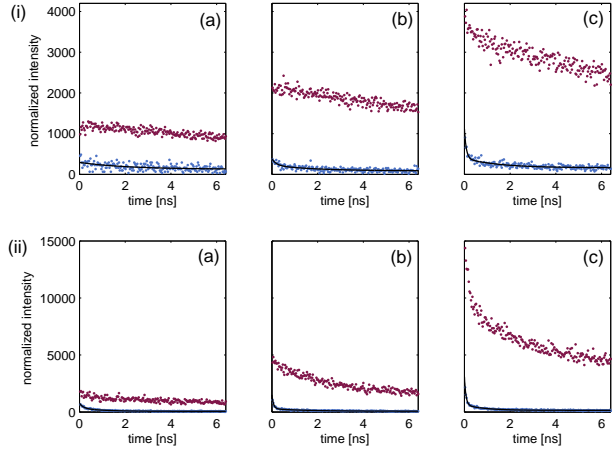}
\caption{`On' (red) and `off' (blue) decay curves taken from two QDs
((i) and (ii)) of 3.8nm diameter at excitation rates of (a) 0.2, (b)
0.4 and (c) 1 photons per QD per pulse. The `off' state decay curves
correspond to $1/e$ decay times of (ia) 1960 $\pm$ 570ps (ib) 660
$\pm$ 260ps (ic) 200 $\pm$ 30ps and (iia) 400 $\pm$ 20ps (iib) 144
$\pm 6$ps (iic) 139 $\pm 4$ps} \label{intensity}
\end{figure}

Further information on the `off' state decay can be extracted from
intensity dependent measurements, performed here on the $3.8$ nm
QDs. For each QD the excitation rate (i.e. the average number of
absorption events per exciton lifetime) was assessed by saturation
of the `on' state emission. In addition, a clear indication of the
average excitation rate approaching unity is the emergence of a fast
multiexciton transient feature in the `on' decay curve. The
measurements of the `off' and `on' decay transients on the same QD
at varying excitation rates reveal a strong correlation between the
excitation intensity and the `off' decay rate. While the effect is
general, it is most dramatically observed in QDs with slow $1/e$
decay times in the `off' state. In Fig.~\ref{intensity} we present
two such examples of `off' and `on' decay transients for single QDs
excited at several intensities, ranging from an average excitation
rate of $\sim 0.2$ to $\sim1$. The two QDs have a low intensity
$1/e$ decay time of $\sim 2$ns (Fig.~\ref{intensity}i(a)) and $\sim
400$ps (Fig.~\ref{intensity}ii(a)). As can be seen, at an elevated
excitation level a fast transient component emerges in the `off'
state decay traces. The multiexponential fit (Fig.~\ref{intensity}c)
gives the fast components lifetimes of (i) $82$ $\pm$ $31$ps and
(ii) $68$ $\pm$ $7$ps, which, when taking into account the $65$ ps
instrument response, correspond to even shorter lifetimes. These
values are in good agreement with the biexciton Auger decay rate of
$\sim 30$ ps for $3.8$nm CdSe QDs \cite{klimov_auger_dots}. The
emergence of such a fast transient at high intensities is observed
in all QDs of this size. In contrast varying the excitation rate at
lower excitation intensities, below $0.1$, revealed no significant
change in the $1/e$ lifetime.

These observations can be summarized as follows: while the decay
rates at low intensities are inconsistent with the assumption of
Auger recombination driven decay, at elevated intensities the Auger
process seems to become the dominant recombination channel.

One cardinal feature of the data, as can be seen in
Fig.~\ref{intensity}, is that in the `off' state the onset of the
fast decay occurs at excitation rates of order $\sim 0.2$, in
contrast to the `on' state, in which the Auger feature emerges when
the excitation rate approaches unity. Based on this key observation,
we wish to offer the following interpretation. The Auger component
in the decay curve indicates the simultaneous presence of more than
one pair of charge carriers. In the `on' state, the extra charge
carriers are provided by multiple excitation of QDs. In the `off'
time, however, the probability of multiple excitation is still low
at the onset of the Auger recombination feature. Therefore, the
additional charge taking part in the Auger process must have a
lifetime longer than the radiative recombination time. On the other
hand, the absence of Auger-like transient at low intensities shows
that the extra charge cannot be present during the entire `off'
period. This implies that either the electron or the hole are
confined to a trap state with a lifetime longer than (but of the
order of) the radiative recombination time, yet orders of magnitude
shorter than the duration of the `off' state. Since no changes are
observed in decay dynamics at excitation rates below $0.1$, we can
estimate the nonradiative recombination time of the trapped charge
in the `off' state as $\approx 10 \tau_{rad}$ ($\approx 200ns$ for
our QDs). The above description assumes that some stochastic process
randomly switches the dot between the `on' state and a range of
`off' states. The physics of this process is responsible for the
observed power law distribution of the `on' and `off' times.

\begin{figure}
\centering
\includegraphics[width=7cm]{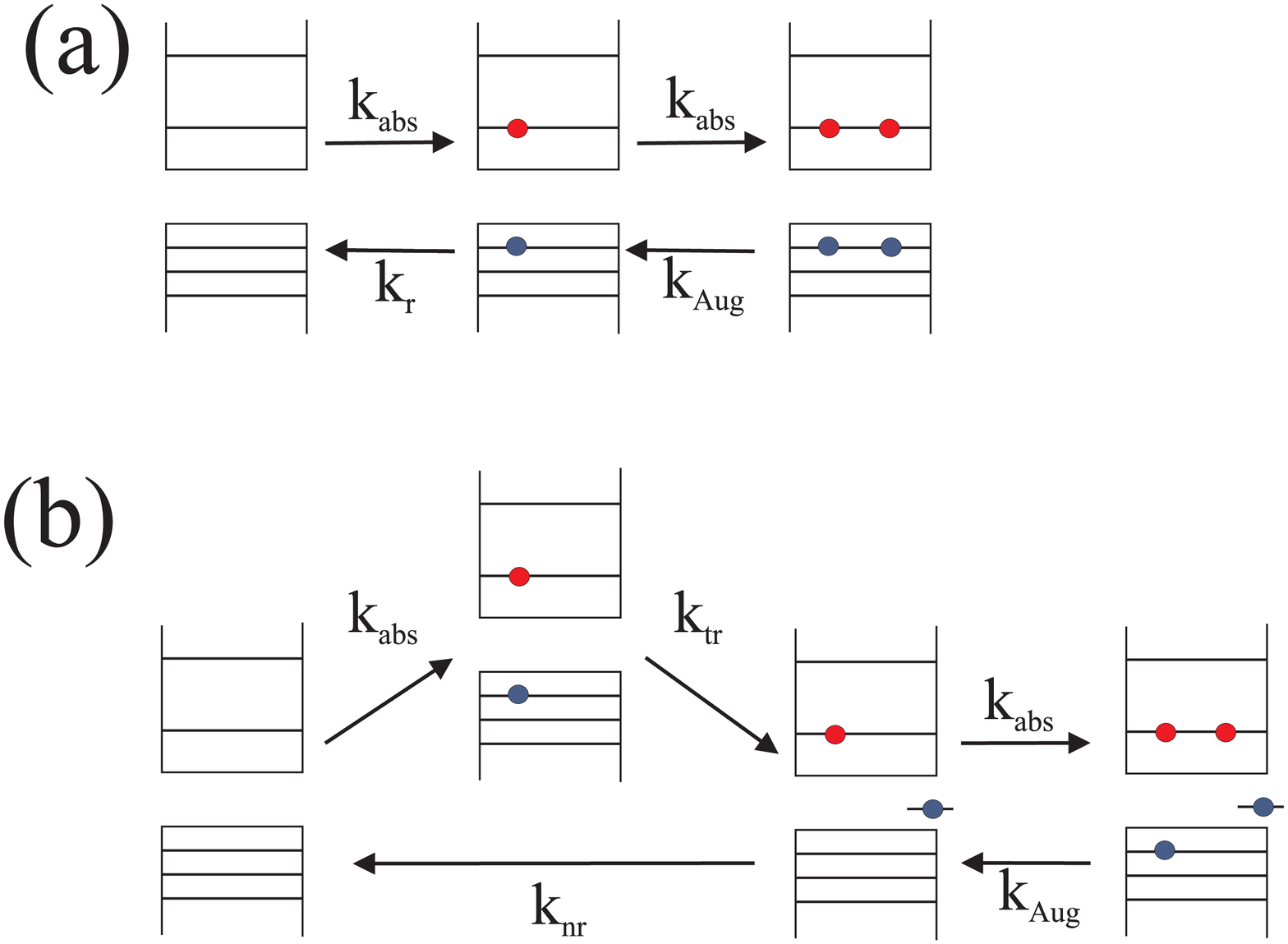}
\caption{Schematic description of the physical processes involved in
(a) `on' and (b) `off' state dynamics. In the `on' state,
predominance of the absorption rate ($k_{abs}$) over the radiative
decay rate ($k_{r}$) results in doubly excited QDs while in the
`off' state the competition is between the absorption rate
($k_{abs}$) and the rate of non-radiative recombination from the
trap ($k_{nr}$).} \label{rates}
\end{figure}

The proposed scheme of QDs charge kinetics is illustrated in
Fig.~\ref{rates}. In the `on' state the dynamics is conventional
(see Fig.~\ref{rates} a); at low intensities the transient dynamics
shows the decay rate of the single exciton $k_r$, while at higher
intensities the Auger recombination with the decay rate $k_{Aug}$
takes over. The above random process `turns off' the QD fluorescence
by opening a transition channel into a trap state either for
electrons or for holes. Thus, rapid trapping at a rate $k_{tr}$,
corresponding to hundreds of picoseconds, inhibits QD luminescence.
At very low intensities, $k_{tr}$ determines the observed `off'
state decay rate and quantum yield. At higher intensities another
exciton can be generated before the trapped charge recombines (with
the rate $k_{nr}$), and Auger recombination becomes visible in the
decay curve (see Fig.~\ref{rates} b). At a low illumination
intensity the `off' QD cycles in a three step loop process including
excitation, rapid trapping of a charge carrier and a relatively long
nonradiative recombination process (left part of the picture in
Fig.~\ref{rates} b). When $k_{abs}$ is increased and becomes
comparable to $k_{nr}$, the QD shifts to the right part of the
scheme in Fig.~\ref{rates} b and the observed decay rate changes to
$k_{Aug}$.
The relations between the decay rates mentioned above determine whether the `off' state
emission is dominated by trapping dynamics or by Auger dynamics at a given excitation
rate. In particular, it is plausible that for some species of QDs the trapped charge
recombination rate $k_{nr}$ is much smaller than for the CdSe/CdS/ZnS dots. This
corresponds to an effectively long lived trap state, meaning that in the `off' state
under typical experimental conditions such QDs would only exhibit the Auger decay
dynamics.

The nature of the physical process responsible for the time dependence of $k_{tr}$ has
yet to be investigated. Regardless of the origin of this process, it is clear that the
above scheme of QD operation is inconsistent with physical models of QD blinking assuming
that a charge must be trapped during the entire `off' period. The experimental results
seem to be consistent, however, with the model proposed by Frantsuzov and Marcus
\cite{frantsuzov}, wherein the random process responsible for the time variation of the
trapping rate is the spectral diffusion of the energy levels in QDs.

Several types of nanocrystals have recently shown nonblinking or nearly nonblinking
behavior \cite{nonblinking_french,nonblinking_klimov,nonblinking_krauss}. significantly
reduced blinking has also been observed by modification of the surrounding matrix of the
QDs \cite{weiss}. Based on the above understanding, efficient elimination of surface
trapping and rapid nonradiative recombination upon trapping are sufficient to eliminate
blinking. Long Auger recombination lifetimes are not required for this to occur, but help
in supporting a high quantum yield despite the existence of traps with relatively slow
nonradiative recombination lifetimes.

In summary, the data presented demonstrates that Auger recombination
alone cannot account for QD blinking. The comparison of the decay
curve intensity dependence for the `on' and `off' states suggests
that the darkening of QDs involves fast trapping of a charge carrier
in a relatively short lived trap state, as opposed to the
conventional idea of the a charge trapped throughout the entire
`off' time. The operation of QDs during the `off' times can
therefore be described as a three-step cyclic process of excitation,
trapping and slow non-radiative relaxation. This phenomenological
model can serve as a basis for the future research on the
microscopic mechanisms of QD blinking.

Financial support by the Minerva foundation and by the Israeli Science Foundation (Grant
No. 1621/07) is gratefully acknowledged.


\begin{thebibliography}{}

\bibitem{Alivisatos_review} A.P. Alivisatos , W.W. Gu , C. Larabell,
Annual Review of Biomedical Engineering {\bf 7}, 55 (2005).

\bibitem{nirmal} M. Nirmal et al., Nature {\bf 383}, 802 (1996).

\bibitem{barkai_review} F.D. Stefani, J.P. Hoogenboom, E. Barkai,
Physics Today 34 (2009).

\bibitem{frantsuzov_review} P. Frantsuzov, M. Kuno, B. Janko, R.A.
Marcus, Nature Phys. {\bf 4}, 519 (2008).

\bibitem{nonblinking_french} B. Mahler et al., Nature Materials {\bf
7}, 659 (2008)

\bibitem{nonblinking_klimov} Y. Chen et al., J. Am Chem. Soc {\bf
130}, 5026 (2008).

\bibitem{nonblinking_krauss} X. Wang et al., Nature {\bf 459}, 686 (2009).

\bibitem{grey} P. Spinicelli et al., Phys. Rev. Lett. {\bf 102},
136801 (2009).

\bibitem{mulvaney} D.E. Gomez, J. van Embden, P. Mulvaney, M.J. Ferne, H. Rubinsztein-Dunlop
ACS Nano {\bf 3}, 2281 (2009).

\bibitem{efros} Al. L. Efros, M. Rosen, Phys. Rev. Lett. {\bf 78},
1110 (1997).

\bibitem{klimov_auger_dots} V.I. Klimov, A.A. Mikhailovsky, D.W. McBranch,
C.A. Leatherdale, M.G. Bawendi, Science {\bf 287}, 1011 (2000).

\bibitem{verberk} R. Verberk, A. M. van Oijen, M. Orrit, Phys. Rev. B {\bf
66}, 233202 (2002).

\bibitem{bawendi_diffusion} K.T. Shimizu et al., Phys. Rev. B {\bf
63}, 205316 (2001).

\bibitem{frantsuzov} P.A. Frantzsuzov, R.A. Marcus, Phys. Rev. B
{\bf 72}, 155321 (2005).

\bibitem{marcus} M. Pelton, G. Smith, N.F. Scherer, R.A. Marcus,
Proc. Natl. Acad. Sci. {\bf 104}, 14249 (2007).

\bibitem{mews_time_res} G. Schlegel, J. Bohnenberger, I. Potapova,
A. Mews, Phys. Rev. Lett. {\bf 88}, 137401 (2002).

\bibitem{bawendi_time_res} B.R. Fisher, H.-J. Eisler, N.E. Scott,
M.G. Bawendi, J. Phys. Chem B {\bf 108}, 143 (2004).

\bibitem{alivisatos_time_res} K. Zhang et al., Nano Lett. {\bf 6},
843 (2006).

\bibitem{peng} Z.A. Peng, X. Peng, J. Am Chem. Soc. {\bf 123}, 183 (2001).

\bibitem{silar} J. Li et al., J. Am. Chem. Soc. {\bf 125}, 12567 (2003).

\bibitem{peng_cs} W. Yu, L. Qu, W. Guo, X. Peng, Chem. Mater. {\bf 15},
2845 (2003).

\bibitem{klimov_auger_rods} H. Htoon, J.A. Hollingsworth, R. Dickerson, V.I.
Klimov, Phys. Rev. Lett. {\bf 91}, 227401 (2003).

\bibitem{trion} P.P. Jha, P. Guyot-Sionnest, ACS Nano {\bf 3}, 1011 (2009).

\bibitem{weiss} H. He et al., Angew. Chem {\bf 45}, 7588 (2006) ;
J. Antelman et al., J. Phys. Chem. C {\bf 113}, 11541 (2009).


\end{thebibliography}
\end{document}